\documentclass[10pt,authoryear]{article}
\usepackage{appendix}
\usepackage{setspace}
\usepackage[font=small,  margin=2cm]{caption}
\usepackage[tbtags]{amsmath}
\usepackage{amsthm,amssymb,amsfonts}
\usepackage[numbers]{natbib}
\usepackage{multirow}
\usepackage{lscape}
\usepackage{subfigure}
\usepackage{makecell}
\usepackage{exscale}
\usepackage{booktabs}
\usepackage{array}
\usepackage{fullpage}
\usepackage{url}
\usepackage{algorithm}
\usepackage{algpseudocode}
\usepackage{bm}
\usepackage{smile}
\usepackage{mathtools}
\usepackage{wrapfig}
\usepackage{lipsum}
\usepackage{mathrsfs}
\usepackage{relsize}
\usepackage{dsfont}
\usepackage{multirow}
\usepackage{apalike}
\usepackage{chngcntr}
\usepackage[usenames,dvipsnames,svgnames,table]{xcolor}
\usepackage[colorlinks=true,
linkcolor=blue,
urlcolor=blue,
citecolor=blue]{hyperref}
\usepackage{ulem}

\newcommand{\bargmin}{\mathop{\mathrm{arg\ min}}}
\numberwithin{equation}{section}
\numberwithin{theorem}{section}
\numberwithin{corollary}{section}
\counterwithout{asmp}{section}
\numberwithin{definition}{section}

\begin{document}

\title{\LARGE  Joint Learning of Multiple Differential Networks with fMRI data for Brain Connectivity Alteration Detection}

	\author{Hao Chen\thanks{Beijing International Center for Mathematical Research, Peking University, Beijing, China},~~Ying Guo\thanks{Department of Biostatistics and Bioinformatics, Rollins School of Public Health,
Emory University, Atlanta, USA; },~~Yong He\thanks{ Shandong University, Jinan, China; Email:{\tt heyong@sdu.edu.cn}. Corresponding Author.},~~Dong Liu\thanks{Shanghai University of Finance and Economics, Shanghai, China;}, ~~ Lei Liu\thanks{Division of Biostatistics, Washington University in St.Louis, St. Louis, USA;}, ~~ Xiaohua Zhou\thanks{Department of Biostatistics, Peking University, Beijing, China; Beijing International Center for Mathematical Research, Peking University, Beijing, China  }}	
	\date{}	
	\maketitle

\textbf{Abstract:} In this study we focus on the problem of joint learning of multiple differential networks with function Magnetic Resonance Imaging (fMRI) data sets from multiple research centers. As the research centers may use different scanners and imaging parameters, joint learning of differential networks with fMRI data from different centers may reflect the underlying mechanism of neurological diseases from different perspectives while capturing the common structures. We transform the task as a penalized logistic regression problem, and exploit sparse group Minimax Concave Penalty (gMCP)  to induce common structures among multiple differential networks and the sparse structures of each differential network. To further enhance the empirical performance, we develop an ensemble-learning procedure. We conduct thorough simulation study to assess the finite-sample performance of the proposed method and compare with  state-of-the-art alternatives. We apply the proposed method to analyze fMRI datasets related with Attention Deficit Hyperactivity Disorder from various research centers. The identified common hub nodes and differential interaction patterns coincides
with the existing experimental studies..

\vspace{0.2em}


\section{Introduction}\label{s:intro}

Attention-Deficit Hyperactivity Disorder (ADHD) is one of the most common neurodevelopmental disorders diagnosed in childhood, which often lasts into adulthood. Children with ADHD may be overly active and have trouble paying attention and controlling impulsive behaviors. Brain connectivity analysis has been at the foreground
of revealing the pathological mechanism of ADHD \citep{Zhu2018multiple,Zhang2020Mixed}.
Accumulated evidence suggests that compared to a healthy brain,
the connectivity network alters in the presence of ADHD \citep{Xia2018matrix,Ji2020Brain}. Extant imaging studies
also heavily support that ADHD involves a
distributed pattern of brain connectivity alterations \citep{Konrad2010Is,Bos2017Structural,Guo2020Shared}.
 Differential network modelling, also known as network comparison, has thus been a powerful tool to capture the alterations of brain connectivity between healthy and diseased groups and reveal the underlying pathological mechanism of  ADHD \citep{zhao2014direct,Xia2015Testing,Yuan2015Differential,tian2016identifying,ji2017jdinac,He2018a,grimes2019integrating,Chen2020Simultaneous}.  In this article, we focus on the problem of comparing brain functional connectivity patterns across
 the diseased group versus the healthy control group, i.e., capturing partial correlations' difference between pairs of brain regions in two groups. The superiority of partial correlation in defining connectivity has become a consensus in the neuroscience community, as it measures the ``direct" connectivity between pairs of regions and thus avoids spurious effects in network modeling, see  \cite{Smith2012The,WangAn2016}.

The functional Magnetic Resonance Imaging (fMRI), which measures changes in blood flow and oxygenation at individual voxels of brain over time, has become the mainstream imaging modalities to study brain functional connectivity. To avoid spurious correlations caused by close spatial proximity, it is a common practice to parcellate
the brain and map brain voxels to a list of pre-specified brain regions of interest (ROIs), then average the time courses of voxels within the same region, which results in a region by time matrix for each fMRI scan.

The example motivating the current work is the ADHD-200 study of resting-state fMRI of children and adolescents, which incorporates demographical information and resting-state fMRI of  both combined types of ADHD and typically developing controls (TDC) from  eight participating sites; for more details, see \cite{neurobureau.projects.nitrc.org}.
Due to the diverse scanners and imaging parameters of the  participating sites, fMRI data from different participating sites may be heterogeneous with distinct statistical properties. However, it is naturally expected that the differential networks between the ADHD and TDC groups from  different participating sites share many common edges, which reflect the intrinsic underlying mechanism of ADHD from different aspects. Estimating a single differential network at each site ignores the partial homogeneity in the differential network structures, which often leads to low power.  It is thus important to develop a joint learning method of  multiple differential networks, which would effectively utilize the shared common structures and consequently increase the estimation precision.

In this work, we propose a Joint Learning method of Multiple Differential Networks (JLMDN) for the spatial$\times$temporal matrix-valued fMRI data.
 Few work exists on joint estimation of multiple differential networks. \cite{Zhang2017Incorporating} and \cite{Ou2018Joint} considered estimating  multiple differential networks for vector-valued data and both modeled the differential network as the difference of the
precision matrices and impose various penalties on its elements. As far as we know, this is the first work on joint estimation of multiple differential networks for matrix-valued fMRI data and the framework is totally different from the work by \cite{Zhang2017Incorporating} and \cite{Ou2018Joint}. We transform the problem of jointly learning multiple differential networks into learning multiple logistic regression models and impose  group penalty on the coefficients to induce common structures of differential networks. We also propose an ensemble learning algorithm to  further enhance the empirical performance of JLMDN.

Our main contributions can be summarized as follows. Firstly, this is the first work on joint estimation of multiple differential networks for matrix-variate fMRI data. Secondly, though we focus on differential network analysis, the proposed  JLMDN can also achieve classification for matrix-variate fMRI data simultaneously. Thirdly, the JLMDN is an ensemble  machine learning procedure and thus more robust and powerful. Finally, the JLMDN can adjust for confounding factors and take the heterogeneity of multiple datasets into account in  network comparison. The advantages of JLMDN are illustrated both through simulation studies and the real fMRI data of ADHD. Although we focus on disease pathologies  of ADHD for illustration in the current work, our proposal also works for other neurological diseases such as Alzheimer's disease.

The rest of this paper is organized as follows. In Section 2, we present the detailed  procedure of JLMDN. In Section  3, we compare the performances of our method with some state-of-the-art competitors via simulation studies. Section 4 illustrates the proposed method through an ADHD study. We summarize our method and give promising future work directions in Section 5.

\section{Preliminaries}
In this section, we introduce the basic matrix-normal framework for differential network analysis.  The following  notations are adopted  throughout the paper. For a real number $x$, let $x_{+}=xI(x\geq 0)$, where $I(\cdot)$ is the indicator function. For any vector $\ba=(a_1,\ldots,a_d)^\top \in \RR^d$, let $\|\ba\|_2=(\sum_{i=1}^d a_i^2)^{1/2}$, $\|\ba\|_1=\sum_{i=1}^d |a_i|$.  { Let $\Ab=[a_{ij}]$ be a square matrix of dimension $d$, $\Ab_{\cdot,j}$ the $j$-th column of $\Ab$, and $\Ab_{i,\cdot}$ the $i$-th row of $\Ab$. Denote $I(\Ab\neq 0)$ as the matrix with the $(i,j)$-th element being $I(a_{ij}\neq 0)$. For a real number $\tau$, denote $I(\Ab > \tau)$ as the matrix with the $(i,j)$-th element being $I(a_{ij}> \tau)$.
Let $\|\Ab\|_1=\sum_{i=1}^d\sum_{j=1}^d |a_{ij}|$, $\|\Ab\|_{\infty}=\max|a_{ij}|$. We denote the trace of $\Ab$ as $\text{Tr}(\Ab)$ and let $\text{Vec}({\Ab})$ be the vector obtained by stacking the columns of $\Ab$. Let $\text{Vec}(b_{ij})_{j>i}$ be the operator that  stacks the columns of the upper triangular elements of
matrix $\Bb=(b_{ij})$ excluding the diagonal elements to a vector. For instance, $\Bb=(b_{ij})_{4\times 4}$, then $\text{Vec}(b_{ij})_{j>i}=(b_{12},b_{13},b_{14},b_{23},b_{24},b_{34})^\top$.
 The notation $\otimes$ represents Kronecker product. For a set $\mathcal{H}$, denote by $\#\{\mathcal{H}\}$ the cardinality of $\mathcal{H}$.

Let $\Xb_{p\times q}$ be the spatial-temporal matrix-valued variate whose rows denote $p$ spatial locations and columns denote $q$ time points. We assume that $\Xb_{p\times q}$ has the matrix-normal distribution with the Kronecker product covariance structure.

\begin{definition}\label{def:1}
 A matrix-variate $\Xb_{p\times q}$ follows the matrix normal distribution
$\Xb_{p\times q}\sim \cN_{p,q}(\Mb_{p\times q},\bSigma_{T}\otimes\bSigma_{S})$,
   if and only if $\text{Vec}(\Xb_{p\times q})$ follows the multivariate normal distribution with mean $\text{Vec}(\Mb_{p\times q})$ and covariance $\bSigma=\bSigma_{T}\otimes\bSigma_{S}$.
\end{definition}

   In Definition \ref{def:1},  $\bSigma_{S}\in \RR^{p\times p}$ and $\bSigma_{T}\in \RR^{q\times q}$ are the covariance matrices of $p$ spatial locations and $q$ times points, respectively. The matrix-normal distribution framework  is widely adopted in neuroimaging analysis and brain connectivity analysis study, see for example, \cite{Yin2012Model,Chenlei2012Sparse,Zhou2014Gemini,Xia2017Hypothesis,Zhu2018multiple,Xia2018matrix,Chen2020Simultaneous}. By assuming that $\Xb_{p\times q}\sim \cN_{p,q}(\Mb_{p\times q},\bSigma_{T}\otimes\bSigma_{S})$, we have
   {\setlength{\abovedisplayskip}{3pt}
   	\setlength{\belowdisplayskip}{3pt}
   \[
   \text{Cov}^{-1}\big(\text{Vec}(\Xb_{p\times q})\big)=\bSigma_{T}^{-1}\otimes\bSigma_{S}^{-1}=\bOmega_{T}\otimes\bOmega_{S},
   \]}and  $\bOmega_{S}\in\RR^{p\times p}$ and $\bOmega_{T}\in\RR^{q\times q}$ denote the spatial precision matrix and temporal precision matrix, respectively. It is obvious that $\bSigma_{S}$ and $\bSigma_{T}$  are only identifiable up to a scaled factor. In the Brain Connected Analysis (BCA), the partial correlation, a.k.a. the scaled precision matrix,  is a commonly adopted correlation measure \citep{Peng2009Partial,Smith2012The,WangAn2016,Zhu2018multiple}. A zero partial correlation coefficient implies the conditional independence of two nodes given all others in the graph when assuming the nodes are jointly normal. In addition, the primary interest in the BCA study is to infer the connectivity network captured by the spatial precision matrix $\bOmega_{S}$.  By contrast, the mean term $\Mb_{p\times q}$ and
the temporal precision matrix $\bOmega_{T}$ are of little interest, which will be treated as nuisance parameters hereafter. In this paper, we simply assume that $\bSigma_{T}=\bOmega_{T}=\Ib_{q\times q}$. Otherwise, the classical technique ``whitening" (\cite{Manjari2016Mixed,Xia2017Hypothesis,Chen2020Simultaneous}) can be adopted to preprocess the data and induce independent columns of each individual matrix observation.
Brain connectivity analysis is in essence to estimate the region-by-region spatial partial correlation matrix $\bR_{S}=\bD_{S}^{-1/2}\bOmega_{S}\bD_{S}^{-1/2}$, where $\bD_{S}$ is the  diagonal matrix of $\bOmega_{S}$.
The advantages of partial correlation over other measures have been recognized in the neuroscience community, as it quantifies the direct connectivity strength between two regions and  avoids spurious effects in network modeling, see  \cite{Smith2012The,WangAn2016,Chen2020Simultaneous,Qiu2021Inference}. In the case and control study, it is of interest to investigate how the network of connected brain regions changes from the case group to the control group, which may provides deep insight into the pathological mechanism of the disease. Typically, it is modelled as the difference of the spatial partial correlation matrices of the case and control groups \citep{Xia2017Hypothesis,Zhu2018multiple,Chen2020Simultaneous}, which is also adopted in the current paper. That is, assume the observations of the ADHD group are from the population $\Xb_{p\times q}\sim \cN_{p,q}(\Mb_{p\times q}^X,(\bOmega_{T}^X\otimes\bOmega_{S}^X)^{-1})$, while those of the control group are from $\Yb_{p\times q}\sim \cN_{p,q}(\Mb_{p\times q}^Y,(\bOmega_{T}^Y\otimes\bOmega_{S}^Y)^{-1})$, we are interested in recovering the support of $\bDelta=\bOmega_{S}^Y-\bOmega_{S}^X$, i.e., $\{(i,j):\Delta_{ij}\neq 0, i<j\}$.

\section{Methodology}
In this section, we introduce the detailed procedures for  Joint Learning of Multiple Differential Networks (JLMDN) method.  The JLMDN contains the
following three  steps. First, construct the individual-specific between-region network measures for all datasets, see part (A) in Figure \ref{fig:workflow}. Second, we adopt the bootstrap strategy  and  for each bootstrapped sample, train  Logistic Regression models with group Minimax Concave Penalty, see part (B) in Figure \ref{fig:workflow}. Finally, ensemble the results from the bootstrap procedure to boost  network
comparison performance, see part (C) in Figure \ref{fig:workflow}.  The details are given in the following subsections.

\begin{figure}[!h]
	\centerline{\includegraphics[width=16cm,height=16cm]{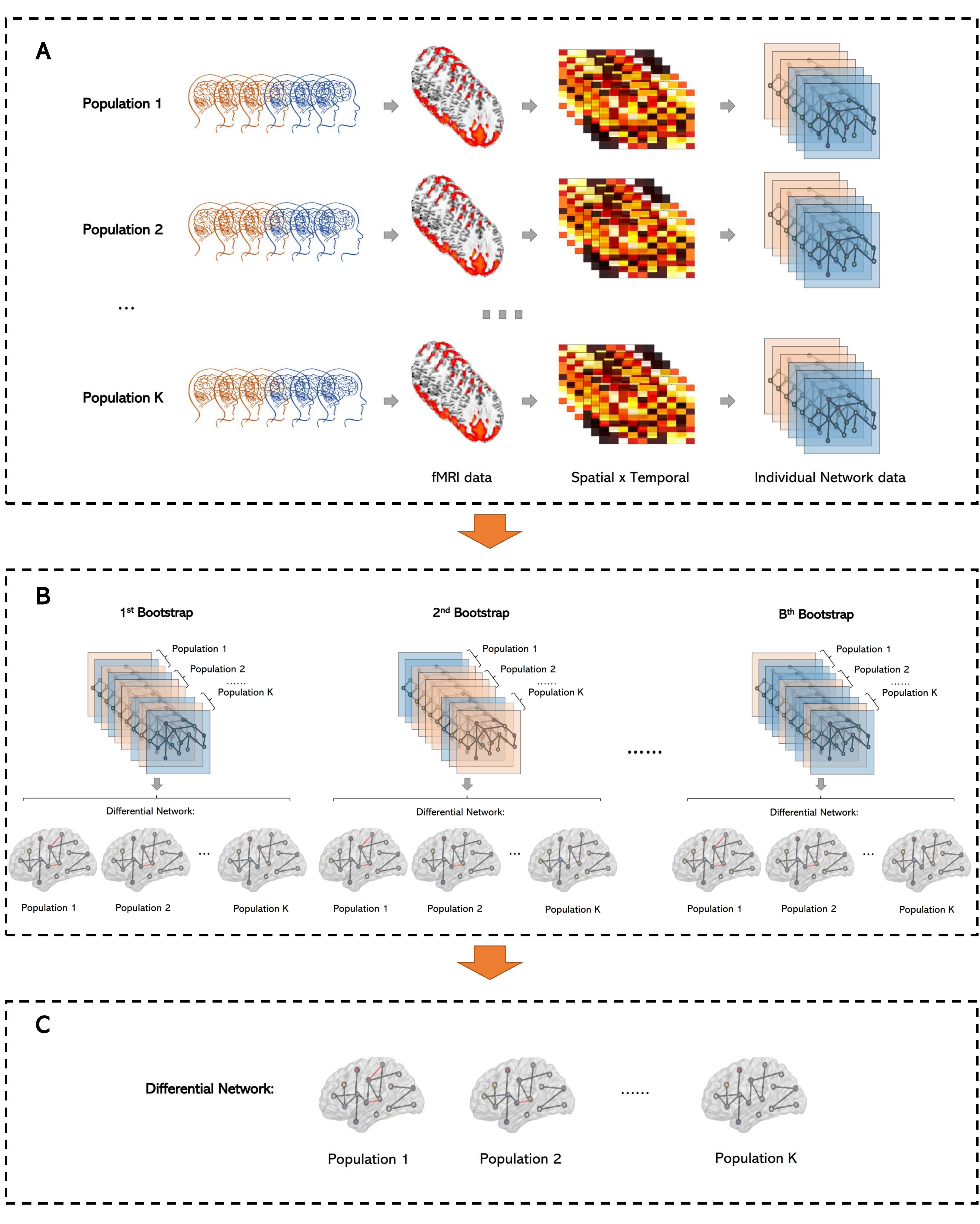}}
	\caption{Workflow of the JLMDN method: (A) construct the individual-specific network strengths with the Spatial $\times$ Temporal data; (B) For each bootstrapped sample, combine individual network information and confounders to fit Logistic Regression  with group Minimax Concave Penalty; (C) Ensemble learning with the bootstrap results.
	}
	\label{fig:workflow}
\end{figure}

\subsection{Individual-specific between-region network measures}
In this section, we introduce how to construct the individual-specific between-region network measures, borrowing idea from \cite{Chen2020Simultaneous}. Assume we have $M$ datasets, and each dataset contains both ADHD and control groups.  We illustrate the procedure with the ADHD group while the control group can be dealt similarly.
Assume that $\Xb^{(\gamma,m)}\in\RR^{p\times q}$ is  the $\gamma$-th subject of the ADHD group in the $m$-th dataset, where the rows denote $p$ spatial locations and columns denote $q$ time points.

The individual spatial sample covariance matrix is given by
\begin{equation}\label{equ:samplecov}
  \hat\bSigma_{S_X}^{(\gamma,m)}=\frac{1}{q-1}\sum_{\nu=1}^q(\bx_{\cdot \nu}^{(\gamma,m)}-\bar\bx^{(\gamma,m)})(\bx_{\cdot \nu}^{(\gamma,m)}-\bar\bx^{(\gamma,m)})^\top,
\end{equation}
where $\bx^{(\gamma,m)}_{\cdot \nu}\in\RR^p$ is the  $\nu$-th column of $\Xb^{(\gamma,m)}$; $\bar\bx^{(\gamma,m)}=1/q\sum_{\nu=1}^q\bx_{\cdot \nu}^{(\gamma,m)}, m=1,\ldots, M$. Note that $p$ is typically larger than or comparable to $q$  for fMRI data.  We  assume that the $\bOmega_{S_X}^{(\gamma,m)}=(\bSigma_{S_X}^{(\gamma,m)})^{-1}$ is sparse, which can be estimated by several high-dimensional penalized methods \citep{friedman2008sparse,Yuan2010High,cai2011constrained}. In consideration of computational efficiency, we adopt  the Constrained
$\ell_1$-minimization for Inverse Matrix Estimation (CLIME) method by \cite{cai2011constrained}, which can be easily implemented by linear programming. In detail,
\begin{equation}\label{equ:clime}
\hat\bOmega_{S_X}^{(\gamma,m)}=\bargmin_{\bOmega}\|\bOmega\|_1, \ \ \text{subject to} \ \ \|\hat\bSigma_{S_X}^{(\gamma,m)}\bOmega-\Ib\|_\infty\leq \lambda,\ \ \bOmega\in\RR^{p\times p},
\end{equation}
where $\lambda$ is a tuning parameter.   The solution of  the optimization problem in (\ref{equ:clime}) is not symmetric in general. We compare the pair of non-diagonal entries at symmetric positions $\hat\Omega_{S_X,ij}^{(\gamma,m)}$ and  $\hat\Omega_{S_X,ji}^{(\gamma,m)}$, and assign the one with  smaller magnitude to both entries.
The tuning parameter $\lambda$ is selected by the {\it Dens} criterion function proposed by \cite{WangAn2016}. We designate the  the desired density level at $0.5$ by setting the parameter ``dens.level" as 0.5 in the function \texttt{DensParcorr} in \textsf{R} package ``DensParcorr". Empirical results show that it works well in various scenarios.

For each individual in the ADHD group of the $m$-th dataset, we estimate the individual spatial precision matrix by the CLIME procedure and obtain $\hat\bOmega_{S_X}^{(\gamma,m)}, \gamma=1,\ldots,n_1^m$ and the scaled version, a.k.a. the partial correlation matrices  $\hat\Rb_{S_X}^{(\gamma,m)}=(\hat\Db_{S_X}^{(\gamma,m)})^{-1/2}\hat\bOmega_{S_X}^{(\gamma,m)}(\hat\Db_{S_X}^{(\gamma,m)})^{-1/2}$,  $\hat\Db_{S_X}^{(\gamma,m)}$ are the diagonal matrix of $\hat\bOmega_{S_X}^{(\gamma,m)}$. Parallelly, denote $\Yb^{(\gamma^\prime,m)}\in\RR^{p\times q}$ as  the $\gamma^\prime$-th subject of the control group in the $m$-th dataset, we obtain $\hat\bOmega_{S_Y}^{(\gamma^\prime,m)}, \hat\Rb_{S_Y}^{(\gamma^\prime,m)} \gamma^\prime=1,\ldots,n_2^m$ for the control group subjects of the $m$-th dataset.  We  define the individual-specific between-region network measures following the strategy of \cite{Chen2020Simultaneous}:
\[
\hat W_{S_X,ij}^{(\gamma,m)}=\frac{1}{2}\log \left(\frac{1+\hat R_{S_X,ij}^{(\gamma,m)}}{1-\hat R_{S_X,ij}^{(\gamma,m)}}\right), \ \ \hat W_{S_Y,ij}^{(\gamma^\prime,m)}=\frac{1}{2}\log \left( \frac{1+\hat R_{S_Y,ij}^{(\gamma^\prime,m)}}{1-\hat R_{S_Y,ij}^{(\gamma\prime,m)}}\right), \ \ 1\leq i<j<p, \ \ 1\leq m\leq M,
\]
where $\hat R_{S_X,ij}^{(\gamma,m)}$ and $\hat R_{S_Y,ij}^{(\gamma^\prime,m)}$ are the $(i,j)$-th element of $\hat\Rb_{S_X}^{(\gamma,m)}$, $\hat\Rb_{S_Y}^{(\gamma^\prime,m)}$, respectively.
Noticeably, the defined between-region network measures are indeed the Fisher transformation of the estimated partial correlations. We adopt the Fisher transformation as it is an approximate variance-stabilizing transformation which alleviates the effects of  underlying skewed distributions of the data and/or the existence of outliers to the following steps.

\subsection{Logistic regression with sparse group Minimax Concave Penalty}
In this section, we introduce how we achieve joint estimation of multiple differential networks by solving a group penalized logistic regression problem.
Let the binary response variable of the $m$-th dataset be denoted as $Z^m$ and its observations be denoted as $\{Z_1^m,\ldots,Z_{n^m}^m\}$, where
\[
Z_k^m=\left\{
\begin{array}{ccc}
1     &      & k\in \text{ADHD};\\
0    &      & k\notin \text{ADHD};
\end{array} \right. \ \  k=1,\ldots,n^m, \ \  n^m=n_1^m+n_2^m \ \ m=1,\ldots,M.
\]
Denote $P^m$ as the probability of the event $Z^m=1$, i.e., $P^m=\Pr(Z^m=1)$. The logistic model for ADHD outcome of the $m$-th dataset is:
\begin{equation}\label{equ:logit}
\text{logit}(P^m)=\log\left(\frac{P^m}{1-P^m}\right)=\sum_{l=1}^L\eta_l^mQ_l^m+\sum_{i=1}^p\sum_{j>i}^p\beta_{ij}^mW^m_{S,ij}, \ \ m=1,\ldots,M,
\end{equation}
where $\bQ^m=(Q_1^m,\ldots,Q_L^m)^\top$ denotes the confounder covariates of the $m$-th dataset (e.g. age and gender) and   $W_{S,ij}^m$ is the Fisher transformation of the spatial partial correlation  between the $i$-th and $j$-th regions $R^{m}_{S,ij}$ of the $m$-th dataset, i.e., $$W_{S,ij}^m=\frac{1}{2}\log\left(\frac{1+R^m_{S,ij}}{1-R^m_{S,ij}}\right).$$
Let $\bbeta^m=\text{Vec}(\beta^m_{ij})_{j>i}$, $\bm{\eta}^m=(\eta_1^m,\ldots,\eta_L^m)^\top$, $\bTheta^{\beta}=(\bbeta^1,\ldots,\bbeta^M)$ and $\bTheta^{\eta}=(\bm{\eta}^1,\ldots,\bm{\eta}^M)^\top$, then
the joint negative log-likelihood function is
\begin{equation}\label{equ:jointlike}
\mathcal{L}\Big({\bTheta^{\eta},\bTheta^{\beta}}\Big)=\frac{1}{N}\sum_{m=1}^M\bigg\{-\sum_{k=1}^{n_m}\Big[Z_k^m\big(\bm{\eta}^{m\top}\bQ^m_k+\bbeta^{m\top} \bW_k^m\big)-\log\Big(1+\exp\big(\bm{\eta}^{m\top}\bQ^m_k+\bbeta^{m\top} \bW^m_k\big)\Big)\Big]\bigg\},
\end{equation}
where $N=\sum_{m=1}^Mn^m$, $\bQ_k^m=(Q^m_{k1},\ldots,Q^m_{kL})^\top$ is the $k$-th observation of $\bQ^m$, and $\bW_k^m$ is the individual-specific network strengths estimated
from the first-stage model, i.e.,
\[
\bW_k^m=\left\{
\begin{array}{cccccccccccccccc}
\text{Vec}(\hat W_{S_X,ij}^{(k,m)})_{j>i}     &      & 1\leq k\leq n_1^m;\\
\text{Vec}(\hat W_{S_Y,ij}^{(k-n_1,m)})_{j>i}    &      & n_1^m+1\leq k\leq n^m.
\end{array} \right.
\]

Assume the coefficient $\beta_{ij}^m\neq 0$, then there exists an  edge  between the  brain regions $i$ and $j$ in the $m$-th differential network, which is used by logistic regression to discriminate the ADHD subjects from the control subjects in the $m$-th dataset. Thus, with the  support set of  $\bTheta^{\beta}$, we can recover the  edges in all $M$ differential networks.
In (\ref{equ:jointlike}), there are total $(L+p(p-1)/2)*M$ variables, which can be very large especially if $p$ is large. We impose a sparse group Minimax Concave Penalty (gMCP) \citep{Liu2013Integrative,Yang2014Concave} on $\bTheta^{\beta}$ in (\ref{equ:jointlike}), which not only induce sparsity of $\bTheta^{\beta}$, but also induce common structures among columns of  $\bTheta^{\beta}$. In this way, the estimated differential networks are separately sparse, and  share many common edges, which reflect the intrinsic underlying mechanism of ADHD from different aspects. The sparse gMCP penalty is defined as
\[
\sum_{l=1}^{d}\rho\left(\|\bTheta^{\beta}_{l,\cdot}\|_2;\sqrt{M}\lambda_1,\gamma_1\right)+\sum_{l=1}^{d}\sum_{m=1 }^M\rho\left(|\Theta^{\beta}_{l,m}|;\lambda_2,\gamma_2\right),
\]
where $d=p(p-1)/2$, $\Theta^{\beta}_{l,m}$ is the $(l,m)$-th element of $\bTheta^{\beta}$ and the function $\rho(t;\lambda,\gamma)$ is the Minimax Concave Penalty (MCP) function \citep{Zhang2010Nearly}  defined as
\[
\rho(t;\lambda,\gamma)=\lambda\int_0^{|t|}\left(1-\frac{x}{\gamma\lambda}\right)_{+}dx.
\]
In the concave penalty function $\rho(t;\lambda,\gamma)$, $\lambda$ is the tuning parameter, $\gamma$ controls the concavity of $\rho$. The MCP enjoys computational simplicity and outstanding empirical performances compared with other nonconvex penalties \citep{Zhang2010Nearly,Liu2013Integrative}. Finally, our estimates of $(\bTheta^{\eta},\bTheta^{\beta})$ are obtained by the following optimization problem,

\begin{equation}\label{equ:optimization}
(\widehat{\bTheta}^{\eta},\widehat{\bTheta}^{\beta})=\bargmin_{(\bTheta^{\eta},\bTheta^{\beta})}\left\{\mathcal{L}\Big(\bTheta^{\eta},\bTheta^{\beta}\Big)+\sum_{l=1}^{d}\rho\left(\|\bTheta_{l,\cdot}^{\beta}\|_2;\sqrt{M}\lambda_1,\gamma_1\right)+\sum_{l=1}^{d}\sum_{m=1 }^M\rho\left(|\Theta_{l,m}^{\beta}|;\lambda_2,\gamma_2\right)\right\},
\end{equation}
which can be solved by a Majorization-Minimization (MM) technique combined with a coordinate descent algorithm. The detailed algorithm is available at the online Web Appendix S1.  Formulation (\ref{equ:optimization}) is motivated by the following considerations. For each edge $(i,j)$, its existence in the $M$
differential networks is represented by a group of regression coefficients, corresponding to each row of $\bTheta^{\beta}$. The first group MCP penalty in (\ref{equ:optimization}) induces common edges of the $M$ differential networks.  The second MCP penalty induces sparsity of each differential network.

In formulation (\ref{equ:optimization}), there exist tuning parameters: $\gamma_1, \gamma_2, \lambda_1$ and $\lambda_2$, where $\gamma_1$ and $\lambda_1$ are for shrinking the estimated coefficients at group level while $\gamma_2$ and $\lambda_2$ for shrinking the estimated coefficients at the within-group level. The estimation accuracy is jointly determined by these  tuning parameters. As suggested in published articles, also in consideration of  computational cost, one can fix $\gamma_1$ and $\gamma_2$ and treat them to be equal, i.e., $\gamma_1=\gamma_2=\gamma$.  In the simulation study, we tried different values for $\gamma$, i.e., $\gamma=\{1, 5, 10, 15\}$, and found that $\gamma=10$ leads to the best performance. However, for different studies,  $\gamma=10$ would not be the universal superiority value and practitioners should search over multiple values of $\gamma$ and choose the satisfactory one. As for $\lambda_1$ and $\lambda_2$,  let $\lambda_2^{max}$ be the smallest $\lambda_2$ which can shrink all coefficients to zero, and let $\lambda_1^{max}(\lambda_2)$ be the smallest $\lambda_1$ under a  fixed $\lambda_2$ for which all coefficients are shrunk to zero. In numerical study, we adopt a two dimensional grid search with 5-fold cross validation to select the tuning parameters $\lambda_1$ and $\lambda_2$. The detailed formulas of $\lambda_2^{max}$ and $\lambda_1^{max}(\lambda_2)$ are given  at the online Web Appendix S2.

\subsection{Ensemble Learning}

In this section, we introduce an  ensemble learning method which further boosts  the network comparison  performance. For the subjects in each dataset,
 we randomly sample the same number of subjects from the ADHD group and the control group separately with replacement,  and then solve the optimization problem in (\ref{equ:optimization}) with the bootstrapped samples. We  conduct the re-sampling technique for $B$ times and
 denote the regression coefficients obtained from each time as $\{\widehat\bTheta_{b}^{\beta},b=1,\ldots,B\}$. We calculate the differential edge weights matrix, defined as $\bPsi=\sum_{b=1}^B I(\bTheta_{b}^{\beta}\neq 0)/B$. {For a pre-specified threshold $\tau$, the support of the differential networks are $\mathcal{F}=\{(s,m):\Psi_{s,m}>\tau, s=1,\ldots d, m=1,\ldots,M\}$, where $\Psi_{s,m}$ is the $(s,m)$-th element of $\bPsi$. In fact, the support of  the $m$-th column of $\bPsi$ corresponds to the estimated support of the $m$-th differential  network, $m=1,\ldots,M$.
 The threshold $\tau$ can be simply taken as $1/2$. } The simulation study in the following section shows that the bootstrap-assisted ensemble learning boosts the  network comparison  performance by a large margin. We summarize the JEMDN in Algorithm \ref{alg:first}.

\begin{algorithm}[H]
	\caption{Ensemble Learning Algorithm for the JLMDN}\label{alg:first}
	{\bf Input:}  $\cD=\Big\{\Xb^{\gamma,m},\gamma=1,\ldots,n_1^m,\Yb^{\gamma^\prime,m},\gamma^\prime=1,\ldots, n_2^m, \bQ_k^m, k=1,\ldots,n^m, m=1,\ldots,M$\Big\}\\
	{\bf Output:} Differential edge weights matrix $\bPsi$
	\begin{algorithmic}[1]
  \Procedure{}{}
		\State Perform the CLIME procedure in (\ref{equ:clime}) for each subject in  $\cD$.

		\State Calculate the network strengths $\bW_k^m$ for subjects in $\cD$.
\For{ $b\leftarrow 1$ {\bf to} $B$}

        \State Sample $\{(\bQ_k^m,\bW_k^m),k=1,\ldots,n^m, m=1,\ldots,M\}$ from $\cD$ with replacement, and obtain bootstrap samples.

        \State Solve the sparse group MCP problem in (\ref{equ:optimization}) with bootstrap samples, and obtain $(\bm{\widehat\Theta}_{b}^{\eta},\bm{\widehat\Theta}_{b}^{\beta})$.

\EndFor

\State Let $\bPsi=\sum_{b=1}^B I(\bTheta_{b}^{\beta}\neq 0)/B$ and recover the differential network edges by $\mathcal{F}=\{(s,v):\Psi_{s,v}>\tau, s=1,\ldots d, v=1,\ldots,M\}$.
  \EndProcedure
	\end{algorithmic}
\end{algorithm}

\section{Simulation Study}
In this section, we conduct simulation study to better gauge performance of the our proposed method JLMDN in comparison with two existing matrix-variate differential network estimation methods. In view of data heterogeneity, we simulate $M$ datasets in each simulation experiment. Each dataset has $n^m=n_1^m+n_2^m=60  (m=1,2,...,M)$ subjects, half for the ADHD group ($n_1^m=30$) and half for the control group ($n_2^m=30$). For each subject, the simulated observation is a $p\times q$ size matrix. The observation matrix for each subject is first generated from matrix normal distribution with  mean zero, i.e., for the ADHD group in the $m$-th dataset, we generate $n_1^m$ independent samples $\Xb^{(\gamma,m)}$ $(\gamma=1,2,\dots,n_1)$ from $\cN_{p,q}\Big(\zero, \bSigma_X^{(\gamma,m)}\Big)$ with $\bSigma_X^{(\gamma,m)}=\bSigma_{T_X}^{(\gamma,m)}\otimes\bSigma_{S_X}^{(\gamma,m)}$; and for the control group, we generate $n_2$ independent samples $\Yb^{({\gamma^\prime,m})}$ $(\gamma^\prime=1,2,\dots,n_2)$ from $\cN_{p,q}\Big(\zero, \bSigma_Y^{(\gamma^\prime,m)}\Big)$ with $\bSigma_Y^{(\gamma^\prime,m)}=\bSigma_{T_Y}^{(\gamma^\prime,m)}\otimes\bSigma_{S_Y}^{(\gamma^\prime,m)}$. In the following we introduce details of the covariance matrices structure for $M$ datasets.

For the temporal covariance matrices in $m$-th dataset: $\bSigma_{T_X}^{(\gamma,m)}=(\sigma_{{T_X},i,j}^{(\gamma,m)})_{q \times q}$ and $\bSigma_{T_Y}^{(\gamma^\prime,m)}=(\sigma_{{T_Y},i,j}^{(\gamma^\prime,m)})_{q \times q}$, the structural type is the Auto-Regressive (AR) correlation, where $\sigma_{{T_X},i,j}^{(\gamma,m)} =0.5^{|i-j|}$ and $\sigma_{{T_Y},i,j}^{(\gamma^\prime,m)} =0.6^{|i-j|}$. For simplicity, we assume that across the $M$ datasets, the temporal covariance matrices of each subject in the same group are exactly the same. In all simulation studies, in order to remove the effect of $\bSigma_{T}$ and induce independent columns, we adopt the common ``whitening" technique. For the details of ``Whitening" technique, one can refer to \cite{Xia2017Hypothesis} and \cite{Chen2020Simultaneous}.

For the spatial covariance matrices, we consider two structures for all subjects across $M$ datasets, which are Hub graph structure and Small-World (SW) graph structure. We resort to \textsf{R} package ``huge" to generate Hub structure with 5 equally-sized and non-overlapping  graph, and employ \textsf{R} package ``rags2ridges" to generate SW structure with 10 starting neighbors and 5\% probability of rewiring. For further details of these two graph structures, one may refer to \cite{zhao2012huge} and \cite{Wieringen2016Ridge}. Then, based on each graph structure, we generate the base adjacency matrices, i.e., the graph structure locating the positions of non-zero elements of matrices $\bOmega_{S_X}^{(m)}$ and $\bOmega_{S_Y}^{(m)}, m=1,\ldots, M$. Then we fill the non-zero positions in matrix $\bOmega_{S_X}^{(m)}$ with random numbers from a uniform distribution with support $[-0.5,-0.3] \cup [0.3,0.5]$. {We select a set  $E_D^{(m)}$ of the elements in $\bOmega_{S_X}^{(m)}$ and change the sign of the element values in $E_D^{(m)}$ to obtain $\bOmega_{S_Y}^{(m)}$. In detail, given a proportion $\rho^{(m)}$, we define the set $E_D^{(m)}$ as $E_D^{(m)}=\{(i,j): i = 1, 2, \ldots , \rho^{(m)}\cdot p,  j = 1, 2, \ldots , \rho^{(m)}\cdot p, i\neq j, \bOmega_{S_X,i,j}^{(m)}\neq 0\}$, where $\bOmega_{S_X,i,j}^{(m)}$ is the $(i,j)$-th element of $\bOmega_{S_X}^{(m)}$}. To ensure that these two matrices are positive definite, we let $\bOmega_{S_X}^{(m)}=\bOmega_{S_X}^{(m)}+(|\lambda_{\min}(\bOmega_{S_X}^{(m)})|+0.5)\Ib$ and $\bOmega_{S_Y}^{(m)}=\bOmega_{S_Y}^{(m)}+(|\lambda_{\min}(\bOmega_{S_Y}^{(m)})|+0.5)\Ib$. The differential edges for the $m$-th dataset are in essence modeled as the support of $\bDelta^{(m)}$, where $\bDelta^{(m)}=\bOmega_{S_X}^{(m)}-\bOmega_{S_Y}^{(m)}$.
In the simulation study, we set $(p,q)=(100,30)$ and $M=3$.  We consider three combinations of $\{(\rho^{(1)}, \rho^{(2)}, \rho^{(3)})\}$: $\{(\rho^{(1)}, \rho^{(2)}, \rho^{(3)})=\{(40\%,40\%,40\%), (40\%,38\%,36\%), (40\%,36\%,32\%)\}$. Note that in the case $\{(\rho^{(1)}, \rho^{(2)}, \rho^{(3)})=(40\%,40\%,40\%)$, all $M$ datasets share the same differential edges and the larger the differences among $\rho^{(1)}$, $\rho^{(2)}$ and $\rho^{(3)}$, the fewer common differential edges across datasets. All the simulation results are based on 100 replications.

We evaluate the performance of differential network estimation accuracy in terms of  true positive rate (TPR), true negative rate (TNR), and true discovery rate (TDR). Let the  true differential matrix be $\bDelta=(\delta_{ij})$  and its estimate be $\hat\bDelta=(\hat\delta_{ij})$ and the TPR, TNR and TDR are defined as follows:
\[
\text{TPR}=\frac{\#\{(i,j):\delta_{ij}\ne 0\cap\hat{\delta}_{ij}\ne 0\}}{\#\{(i,j):\delta_{ij}\ne 0\}}, \ \ \
\text{TNR}=\frac{\#\{(i,j):\delta_{ij}= 0\cap\hat{\delta}_{ij}= 0\}}{\#\{(i,j):\delta_{ij}= 0\}},
\]
\[
\text{TDR}=\frac{\#\{(i,j):\delta_{ij}\ne 0\cap\hat{\delta}_{ij}\ne 0\}}{\#\{(i,j):\hat{\delta}_{ij}\ne 0\}},
\]
in which the  TPR is also typically referred to as Recall rate and the TDR is also known as Precision rate.
We compare the proposed method JLMDN with two existing  state-of-the-art matrix-variate differential network estimation methods.  The first one is the joint estimation method of multiple graphs for matrix-variate data proposed by \cite{Zhu2018multiple}, denoted as ``Non-Convex". The other is the differential network estimation method for matrix-variate data proposed by \cite{Chen2020Simultaneous} which can also achieve classification  simultaneously, named as ``SDNCMV". As both ``Non-Convex" and  ``SDNCMV" cannot achieve joint differential network estimation for multiple datasets, we adopt each method to do the differential network analysis on  $M$ datasets one by one.

\begin{table}[!p]\caption{{TPR, TNR and TDR averaged over 100 replications (\%) for each dataset with Hub Structure}}\label{tab:hub}
\centering
\scalebox{1}{
\renewcommand{\arraystretch}{1.2}
\begin{tabular*}{16cm}{cclllllllllllll}
\toprule[2pt]
 & &     \multicolumn{3}{c}{Dataset 1} & & \multicolumn{3}{c}{Dataset 2}  & &   \multicolumn{3}{c}{Dataset 3} \\
  \cline{3-5}  \cline{7-9}  \cline{11-13}
  Methods && TPR  & TNR & TDR & & TPR   & TNR &TDR & & TPR   & TNR &TDR  \\
   \hline
   & & \multicolumn{11}{c}{$\bm{\{(\rho^{(1)}, \rho^{(2)}, \rho^{(3)})\}=\{(40\%,40\%,40\%)\}}$}\\

   JLMDN&         &81.0  &100 &94.9 &   &81.0 & 100 &94.7  &       &81.1 &100 &94.6 \\
   SDNCMV&     &64.8 &99.9 &85.2 &      &55.7 &99.9   &86.6  &       &57.2 &99.9 &83.3 \\
   Non-Convex& &99.3 &18.7 &1.5 &    &99.9 & 12.2 &0.9  &       &99.9 &19.9 &1.0 \\
      & & \multicolumn{11}{c}{$\bm{\{(\rho^{(1)}, \rho^{(2)}, \rho^{(3)})\}=\{(40\%,38\%,36\%)\}}$}\\

   JLMDN&        &75.1  &99.9 &96.1 &   &78.9 & 100 &95.6  &       &80.6 &100 &92.5 \\
   SDNCMV&         &69.1 &99.9 &81.3 &    &58.3 & 99.9 &84.1  &       &52.6 &99.9 &86.1 \\
   Non-Convex&    &99.9 &15.7 &0.9 &      &100 &16.6   &0.7  &       &99.9 &19.9 &0.9 \\
         & & \multicolumn{11}{c}{$\bm{\{(\rho^{(1)}, \rho^{(2)}, \rho^{(3)})\}=\{(40\%,36\%,32\%)\}}$}\\

   JLMDN&        &70.6  &100 &94.3 &   &77.6 & 100 &93.1  &       &82.6 &99.9 &88.1 \\
   SDNCMV&         &68.1 &99.9 &84.2 &    &54.5 & 99.9 &87.3  &       &51.6 &99.9 &84.4 \\
   Non-Convex&    &99.9 &20.0 &1.0 &      &99.9 &12.1   &0.8  &       &99.9 &19.8 &0.8 \\

\bottomrule[2pt]
\end{tabular*}}
\end{table}

\begin{table}[!p]\caption{{TPR, TNR and TDR averaged over 100 replications (\%) for each dataset with SW structure}}\label{tab:sw}
\centering
\scalebox{1}{
\renewcommand{\arraystretch}{1.2}
\begin{tabular*}{16cm}{cclllllllllllll}
\toprule[2pt]
 & &     \multicolumn{3}{c}{Dataset 1} & & \multicolumn{3}{c}{Dataset 2}  & &   \multicolumn{3}{c}{Dataset 3} \\
  \cline{3-5}  \cline{7-9}  \cline{11-13}
  Methods && TPR  & TNR & TDR & & TPR   & TNR &TDR & & TPR   & TNR &TDR  \\
   \hline
   & & \multicolumn{11}{c}{$\bm{\{(\rho^{(1)}, \rho^{(2)}, \rho^{(3)})\}=\{(40\%,40\%,40\%)\}}$}\\

   JLMDN&         &62.9  &99.9 &97.8 &   &62.8 & 99.9 &97.9  &       &69.9 &100 &98.1 \\
   SDNCMV&     &61.6 &99.5 &84.6 &      &62.3 &99.7   &89.8  &       &56.1 &99.6 &86.5 \\
   Non-Convex& &97.3 &56.0 &7.7 &    &90.9 & 79.7 &14.4  &       &97.5 &47.6 &6.6 \\
      & & \multicolumn{11}{c}{$\bm{\{(\rho^{(1)}, \rho^{(2)}, \rho^{(3)})\}=\{(40\%,38\%,36\%)\}}$}\\

   JLMDN&        &58.6  &99.9 &97.4 &   &65.0 & 99.9 &97.2  &       &66.8 &99.9 &93.6 \\
   SDNCMV&         &65.0 &99.3 &79.6 &    &65.6 & 99.6 &86.9  &       &59.9 &99.5 &78.7 \\
   Non-Convex&    &97.9 &47.6 &6.6 &      &97.6 &54.6   &6.8  &       &99.6 &14.9 &3.5 \\
         & & \multicolumn{11}{c}{$\bm{\{(\rho^{(1)}, \rho^{(2)}, \rho^{(3)})\}=\{(40\%,36\%,32\%)\}}$}\\

   JLMDN&        &54.9  &99.9 &96.6 &   &65.6 & 99.9 &95.8  &       &68.9 &99.8 &92.0 \\
   SDNCMV&         &63.3 &99.5 &83.2 &    &63.5 & 99.7 &86.8  &       &55.5 &99.6 &83.9 \\
   Non-Convex&    &96.9 &55.6 &7.6 &      &97.7 &54.7   &6.3  &       &99.6 &19.2 &3.3 \\
   \bottomrule[2pt]
\end{tabular*}}
\end{table}

Table \ref{tab:hub} and Table \ref{tab:sw} show the TPRs, TNRs and TDRs by different methods averaged over 100 replications for each dataset with Hub structure and Small-World structure respectively, with the threshold $\tau$ selected by maximizing the sum of TPR and
TDR. Figure \ref{fig:hub} and Figure \ref{fig:sw} depict the Precision-Recall curves for each dataset with different spatial structures. From Table \ref{tab:hub}, when the spatial structure of each dataset is of the Hub type, it can be seen that three methods have comparable TNRs, while JLMDN has comparable TPRs with the Non-Convex method, which are higher  than those of SDNCMV. In terms of TDRs, JLMDN performs uniformly better than the SDNCMV and the Non-Convex methods. Figure \ref{fig:hub} shows the Precision-Recall curves for each dataset with hub structure, which intuitively show that JLMDN outperforms than SDNCMV and Non-Convex as the combination of $(\rho^{(1)},\rho^{(2)},\rho^{(3)})$ varies. Table \ref{tab:sw} and Figure \ref{fig:sw} show the simulation results when the spatial structure of each dataset is of Small-World type, from which we can draw  similar conclusions as for the Hub structure. Most importantly, Figure \ref{fig:hub} and Figure \ref{fig:sw} show that the Precision-Recall curves of JLMDN are similar  across different datasets as the combination of $(\rho^{(1)},\rho^{(2)},\rho^{(3)})$ varies, and the smaller the difference among $\rho^{(1)}$, $\rho^{(2)}$ and $\rho^{(3)}$, the better JLMDN performs than SDNCMV and Non-Convex, which indicates that the JLMDN can exploit the similarity between the true differential networks and improve the estimation performances by a large margin.

\begin{figure}[H]
	\centerline{\includegraphics[width=16cm,height=14cm]{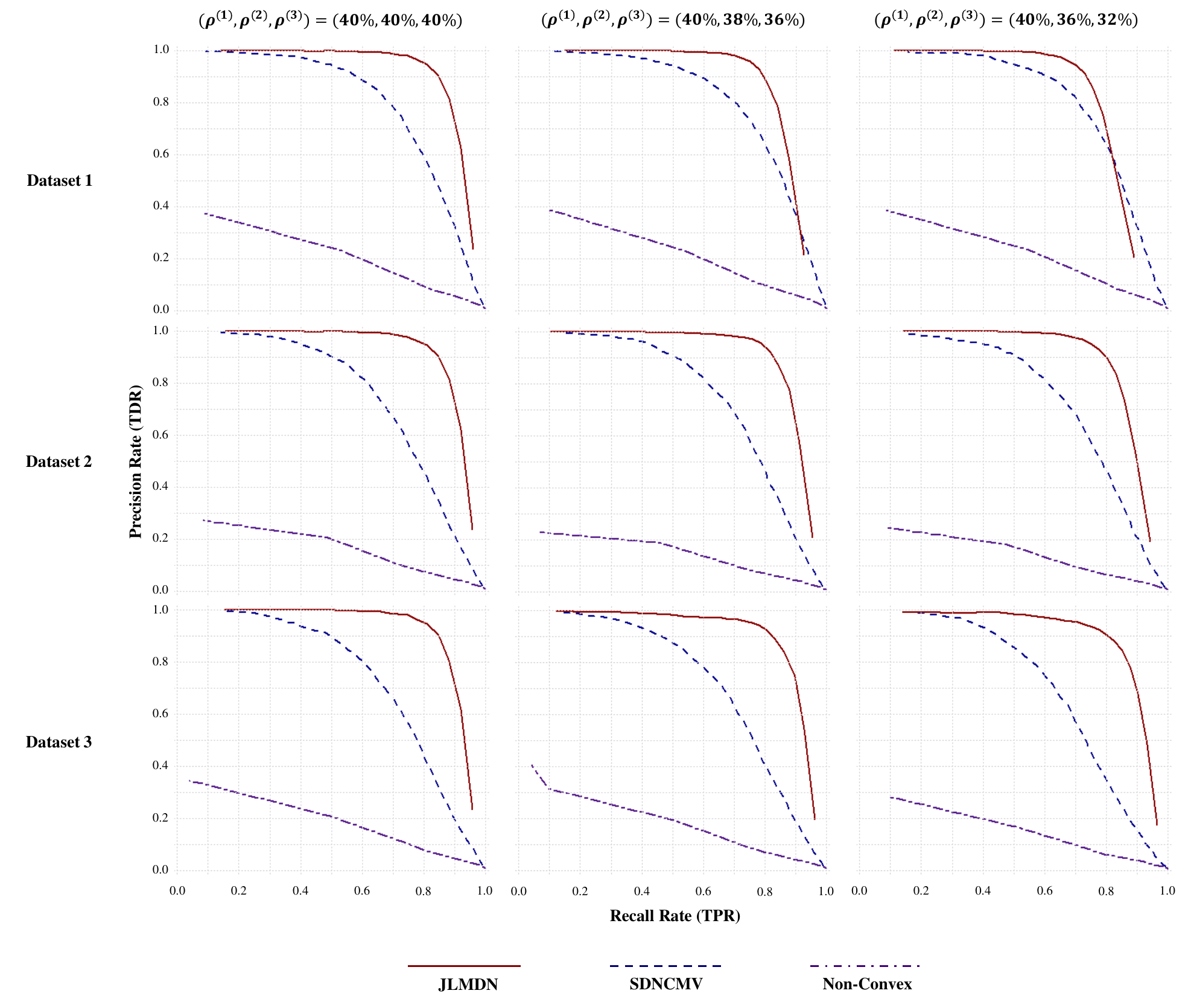}}
	\caption{Precision-Recall curve for each dataset with Hub structure under different combinations of $(\rho^{(1)}, \rho^{(2)}, \rho^{(3)})$.
	}
	\label{fig:hub}
\end{figure}

\begin{figure}[H]
	\centerline{\includegraphics[width=16cm,height=14cm]{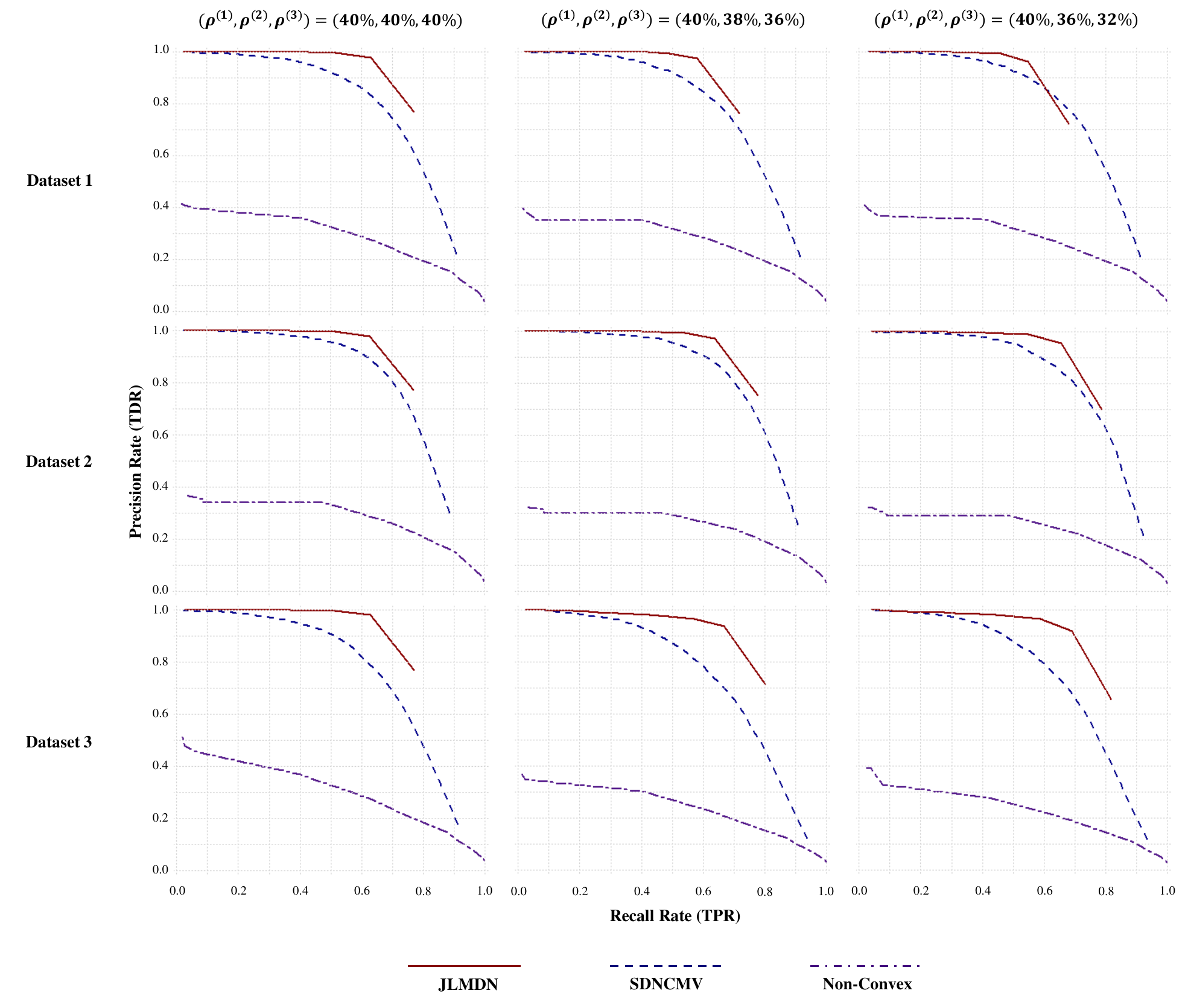}}
	\caption{Precision-Recall curve for each dataset with SW structure under different combinations of $(\rho^{(1)}, \rho^{(2)}, \rho^{(3)})$.
	}
	\label{fig:sw}
\end{figure}
\section{The fMRI Data of Attention Deficit and Hyperactivity Disorder}
 In this section we focus on  differential network analysis for ADHD-200 study of resting-state fMRI of children and adolescents. As mentioned in the Introduction, the ADHD-200 study incorporates demographical
information and resting-state fMRI of  both combined types
of ADHD and typically developing controls (TDC) from  eight participating sites, see \cite{neurobureau.projects.nitrc.org}. We selected two sites from the eight participating sites, the Peking University site and the New York University site. The two datasets that we analyzed are the preprocessed version using the Athena pipeline (\cite{Bellec2017The}). All fMRI scans have been preprocessed by standard procedures, such as slice timing correction, motion correction, spatial smoothing, de-noising by regressing out motion parameters, and white matter and cerebrospinal fluid time courses. For each dataset, the brain images of each individual were mapped into 116 ROIs using the Anatomical Automatic Labeling (AAL) atlas.  We only used the scans that passed the quality control following the strategy in \cite{Xia2018matrix} and \cite{Ji2020Brain}, and the information is provided in the attached phenotypic data of ADHD-200 study. In the case that both scans of a subject passed the quality control, we chose the first scan. In the case that neither scan passed the quality control, we directly deleted this individual from further analysis. The individuals with missing diagnostic status or missing scans were also removed from further analysis, and this finally led to the dataset from Peking University site consisting of 74 combined ADHD individuals and 109 TDC individuals, the dataset from New York University site consisting of 96 combined ADHD individuals and 91 TDC individuals. We averaged the time series of voxels within the same ROI and obtain a spatial-temporal data matrix for each individual collected from Peking University site, with spatial dimension $p=116$, temporal dimension $q=232$, while $p=116$,  $q=172$ for New York University site.

Moreover, as there are total $p*(p-1)/2=6670$ edge variables included in each group $m$ for the method JLMDN,  we first adopt a Sure Independence Screening (SIS) procedure  to alleviate computation burden and improve the identification accuracy of differential edges. We employed the R package ``SIS" to filter out 5670 variables and kept the remaining 1000 variables in the active set. One may refer to \cite{fan2008sis} for more details about SIS. For a fair comparison, we adopt the same SIS procedure for method SDNCMV. The Non-Convex method is not regression-based, and thus we do not employ the SIS procedure.

We assign the 116 ROIs to 8 functional modules. In Figure \ref{fig:bj} and Figure \ref{fig:ny}, we show the top 20 important differential edges identified by different methods from the Beijing Site and New York site, respectively. The ROIs with the same color  belong to the same function modules. All the brain visualizations in this article were created using BrainNet Viewer proposed by \cite{Xia2013BrainNet}. As we can see from Figure \ref{fig:bj} and Figure \ref{fig:ny}, the  brain regions involved in differential edges identified by our method JLMDN are mildly different between the two  sites, while those identified by method SDNCMV are exactly the same and those identified by Non-Convex are very different between the two sites. Clinically, most of the pathogenic brain regions of ADHD should be the same with little difference between the two sites, which reflect the intrinsic underlying mechanism of ADHD  while allowing for heterogeneity between geographical sites. Our method is consistent with the clinical knowledge.

No gold standard is available for evaluating the accuracy of recovering differential edges in the real example since the true underlying pathogenic mechanism is unknown. However, we find evidence for the involved brain regions identified by JLMDN. These  regions  are mainly located in the Frontal, Parietal, SCGM and Cerebellum for the individuals from Peking University site and mainly located in the Frontal, Cerebellum, Parietal and Limbic for the individuals from New York University site. Frontal part is the biggest part in human brain, and it is devoted to action of skeletal movement, ocular movement, speech control, and the expression of emotions.\cite{vaidya2012neuro} has shown that the frontal is important for behavior requiring executive control, an area of impairment in ADHD. \cite{Murias2006Functional} also used electroencephalography (EEG) coherence to imply that altered functional connectivity, particularly among Frontal region, is implicated in ADHD. In addition, the Cerebellum region is also identified as a critical brain region in ADHD pathophysiology by  JLMDN. In fact, the Cerebellum region is involved in multiple functions, including maintenance of body balance, coordination of movements and motor learning. \cite{Stoodley2016cerebellum} pointed out that  the Cerebellum region affects the structure and function of cerebro-cerebellar circuits, which involves the skill acquisition in multiple domains.  When this process is abnormal, it may have some significant impacts on behavior. Cerebellar dysfunction is evident in ADHD. In structural neuroimaging field, \cite{Valera2007meta} found that ADHD is characterized by the abnormalities in Cerebellum region, confirming by a large scale case-controlled study.

\begin{figure}[H]
 \centerline{\includegraphics[width=16cm,height=15.6cm]{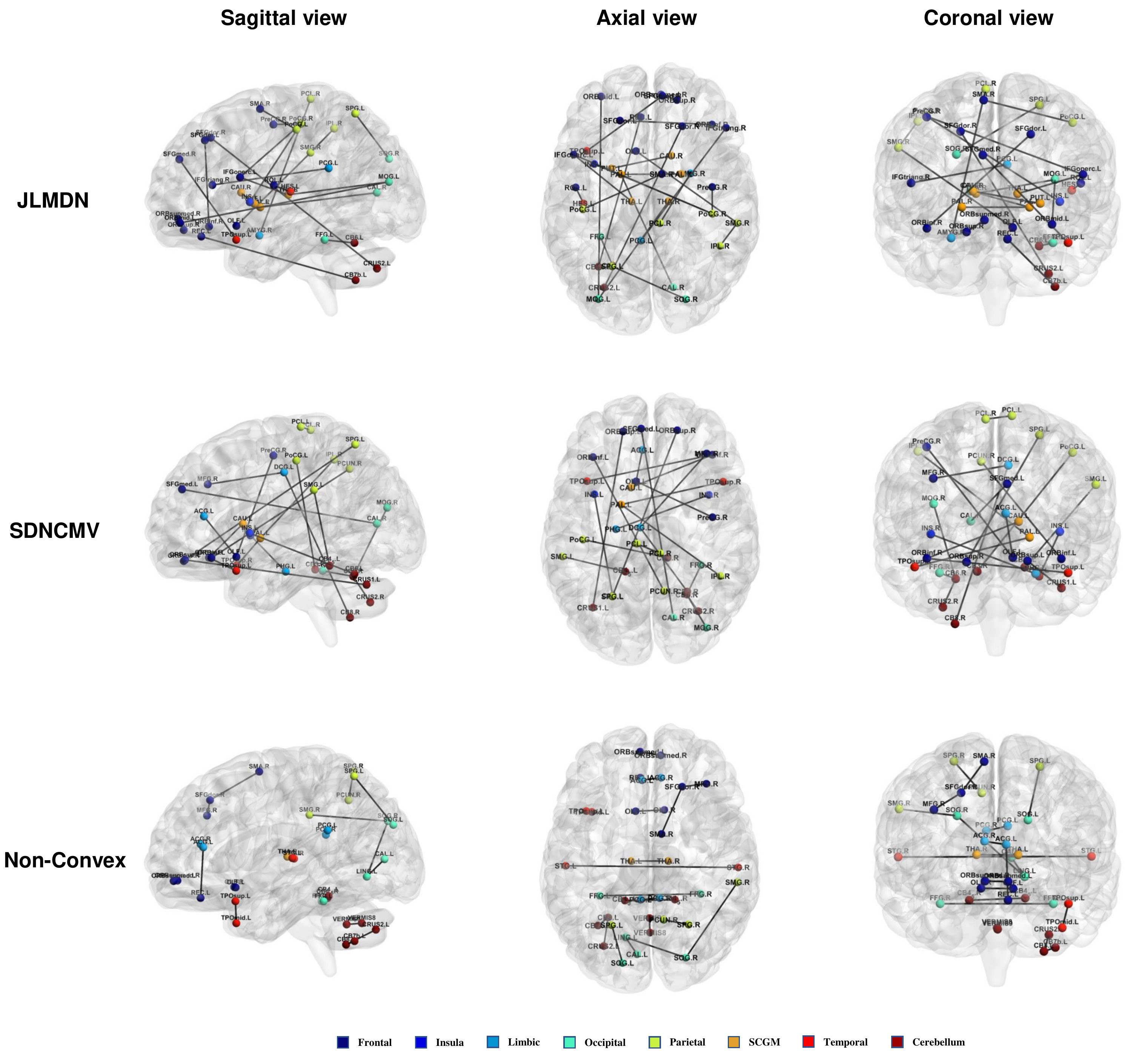}}
 \caption{Differential edges and the associated brain regions identified by various procedures for the ADHD resting-state fMRI data from Peking University site.
 }
 \label{fig:bj}
\end{figure}
\begin{figure}[H]
 \centerline{\includegraphics[width=16cm,height=15.6cm]{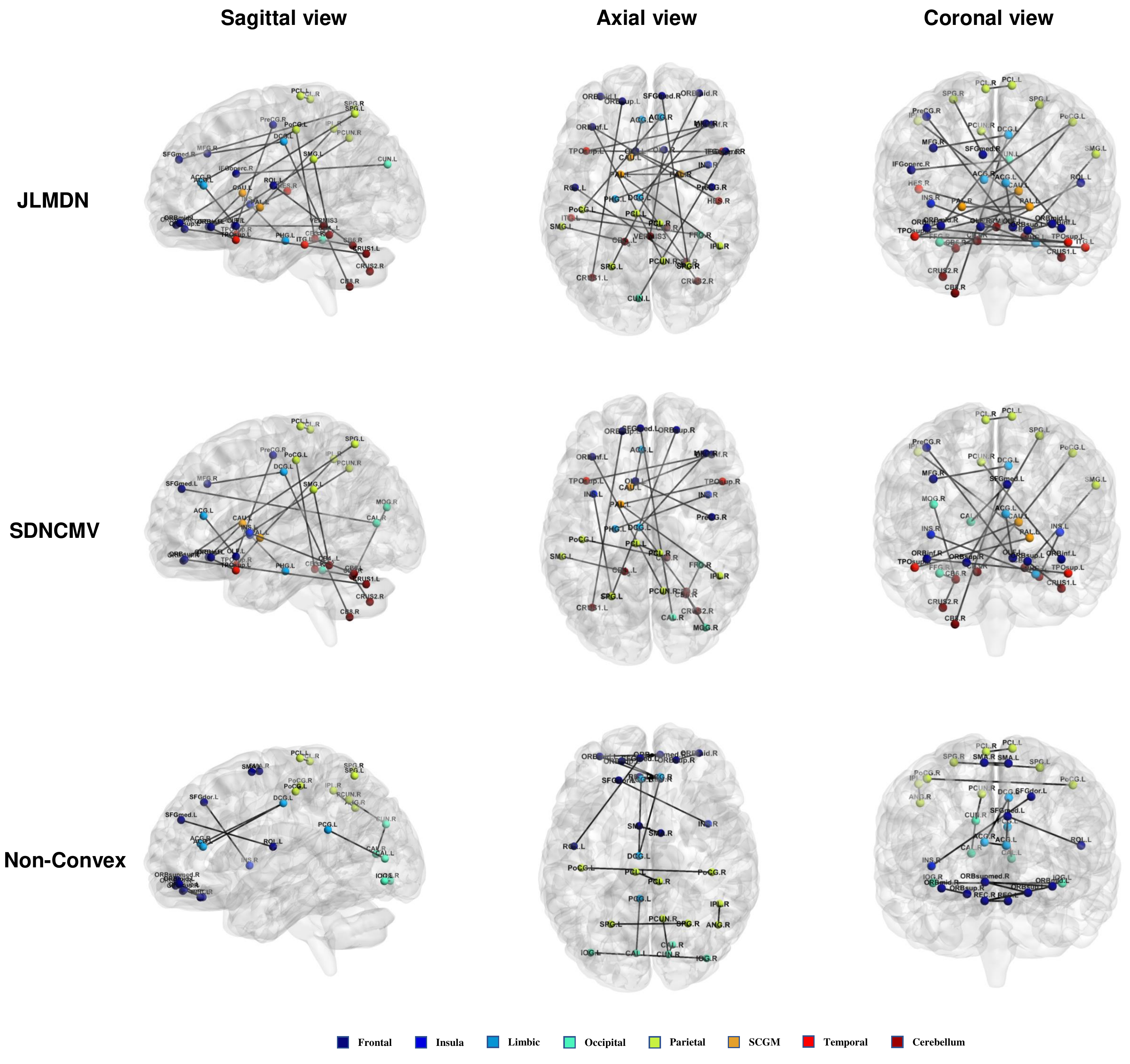}}
 \caption{Differential edges and the associated brain regions identified by various procedures for the ADHD resting-state fMRI data from New York University site.
 }
 \label{fig:ny}
\end{figure}

As it is difficult to objectively evaluate the accuracy of recovering differential edges, we evaluate the classification performance of our method JLMDN, which may also provide an indirect evaluation of accuracy in terms of recovering the differential edges. In particular, it is expected that if a set of differential edge variables has better prediction performance in real dataset, the  corresponding differential edges are more reliable. We employed a resampling approach. At first, we split the dataset from each site into training and testing sets, with sizes 80\%:20\%. Then we combined the training sets of two sites as the ultimate training sets and combined two test sets as the ultimate test sets. We generated classification models using the training set, and predict the class label for each individual in the testing set. We repeat the process  100 times and the average classification error is  0.0\% (standard deviation 0.00). In other word, our  JLMDN method has excellent classification performance. Note that SDNCMV can also do classification while the Non-Convex method cannot. For the dataset from Beijing site, the average classification error over 100 replications by SDNCMV is 7.8\% (standard deviation 0.05) while for the dataset from New York site, the average classification error over 100 replications by SDNCMV is 3.2\% (standard deviation 0.03), which further shows  the superiority of our  JLMDN method over SDNCMV.

\section{Discussion}
In empirical studies, the analysis of a single dataset may lead to unsatisfactory results possibly due to the small sample size.  Integrating multiple datasets would effectively increase sample size and thus lead to more accurate/ reliable results. In the current paper, we focus on estimating the differential network between ADHD patients and normal controls, which would provide deeper insights into the pathology of ADHD.  The collected fMRI datasets from different participating sites  show heterogeneity due to the diverse scanners and imaging parameters of the  participating sites.  We proposed a way to jointly estimate multiple differential networks - JLMDN. The advantage of JLMDN lies in its ability to discover a common structure and jointly estimate common links across differential networks, which leads to improvements thanks to the borrowed information from other related networks. In essence, we transform the problem into solving a sparse group penalized logistic regression model. Thorough numerical studies show the superiority of JLMDN over the methods estimating each differential network separately. We also illustrate the practical utility of the JLMDN method by analyzing the fMRI data of ADHD.  For the optimization problem in (\ref{equ:optimization}), we may further add a  hierarchical group bridge penalty \citep{Zhang2017Incorporating} on $\bTheta^{\beta}$ to induce hub nodes on the common differential network structures. The optimization problem is more challenging and of independent interest as our future work.

\section*{Acknowledgements}
 This work was supported by grants from  the National Natural Science
Foundation of China (Grant No. 11801316, 12026606); Natural Science Foundation of Shandong Province (Grant No. ZR2019QA002); the Fundamental Research Funds of Shandong University  and National Center for Advancing Translational Sciences (Grant No. UL1TR002345 to L.L).

\bibliographystyle{model2-names}
\bibliography{ref}

\begin{thebibliography}{42}
\expandafter\ifx\csname natexlab\endcsname\relax\def\natexlab#1{#1}\fi
\providecommand{\url}[1]{\texttt{#1}}
\providecommand{\href}[2]{#2}
\providecommand{\path}[1]{#1}
\providecommand{\DOIprefix}{doi:}
\providecommand{\ArXivprefix}{arXiv:}
\providecommand{\URLprefix}{URL: }
\providecommand{\Pubmedprefix}{pmid:}
\providecommand{\doi}[1]{\href{http://dx.doi.org/#1}{\path{#1}}}
\providecommand{\Pubmed}[1]{\href{pmid:#1}{\path{#1}}}
\providecommand{\bibinfo}[2]{#2}
\ifx\xfnm\relax \def\xfnm[#1]{\unskip,\space#1}\fi
\bibitem[{Bellec et~al.(2017)Bellec, Chu, Chouinard-Decorte
  et~al.}]{Bellec2017The}
\bibinfo{author}{Bellec, P.}, \bibinfo{author}{Chu, C.},
  \bibinfo{author}{Chouinard-Decorte, F.}, et~al., \bibinfo{year}{2017}.
\newblock \bibinfo{title}{The neuro bureau adhd-200 preprocessed repository}.
\newblock \bibinfo{journal}{Neuroimage} \bibinfo{volume}{144},
  \bibinfo{pages}{275}.
\bibitem[{Bos et~al.(2017)Bos, Oranje, Achterberg, Vlaskamp and
  Durston}]{Bos2017Structural}
\bibinfo{author}{Bos, D.J.}, \bibinfo{author}{Oranje, B.},
  \bibinfo{author}{Achterberg, M.}, \bibinfo{author}{Vlaskamp, C.},
  \bibinfo{author}{Durston, S.}, \bibinfo{year}{2017}.
\newblock \bibinfo{title}{Structural and functional connectivity in children
  and adolescents with and without attention deficit/hyperactivity disorder}.
\newblock \bibinfo{journal}{Journal of Child Psychology $\&$ Psychiatry}
  \bibinfo{volume}{58}, \bibinfo{pages}{810--818}.
\bibitem[{Cai et~al.(2011)Cai, Liu and Luo}]{cai2011constrained}
\bibinfo{author}{Cai, T.}, \bibinfo{author}{Liu, W.}, \bibinfo{author}{Luo,
  X.}, \bibinfo{year}{2011}.
\newblock \bibinfo{title}{A constrained $\ell_1$ minimization approach to
  sparse precision matrix estimation}.
\newblock \bibinfo{journal}{Journal of the American Statistical Association}
  \bibinfo{volume}{106}, \bibinfo{pages}{594--607}.
\bibitem[{Chen et~al.(2021)Chen, Guo, He, Ji, Liu, Shi, Wang, Yu and
  Zhang}]{Chen2020Simultaneous}
\bibinfo{author}{Chen, H.}, \bibinfo{author}{Guo, Y.}, \bibinfo{author}{He,
  Y.}, \bibinfo{author}{Ji, J.}, \bibinfo{author}{Liu, L.},
  \bibinfo{author}{Shi, Y.}, \bibinfo{author}{Wang, Y.}, \bibinfo{author}{Yu,
  L.}, \bibinfo{author}{Zhang, X.}, \bibinfo{year}{2021}.
\newblock \bibinfo{title}{Simultaneous differential network analysis and
  classification for matrix-variate data with application to brain
  connectivity}.
\newblock \bibinfo{journal}{Biostatistics, in press} .
\bibitem[{Fan and Lv(2008)}]{fan2008sis}
\bibinfo{author}{Fan, J.}, \bibinfo{author}{Lv, J.}, \bibinfo{year}{2008}.
\newblock \bibinfo{title}{Sure independence screening for ultrahigh dimensional
  feature space}.
\newblock \bibinfo{journal}{Journal of the Royal Statistical Society: Series B
  (Statistical Methodology)} \bibinfo{volume}{70}, \bibinfo{pages}{849--911}.
\newblock \DOIprefix\doi{https://doi.org/10.1111/j.1467-9868.2008.00674.x}.
\bibitem[{Friedman et~al.(2008)Friedman, Hastie and
  Tibshirani}]{friedman2008sparse}
\bibinfo{author}{Friedman, J.}, \bibinfo{author}{Hastie, T.},
  \bibinfo{author}{Tibshirani, R.}, \bibinfo{year}{2008}.
\newblock \bibinfo{title}{Sparse inverse covariance estimation with the
  graphical lasso}.
\newblock \bibinfo{journal}{Biostatistics} \bibinfo{volume}{9},
  \bibinfo{pages}{432--441}.
\bibitem[{Grimes et~al.(2019)Grimes, Potter and Datta}]{grimes2019integrating}
\bibinfo{author}{Grimes, T.}, \bibinfo{author}{Potter, S.S.},
  \bibinfo{author}{Datta, S.}, \bibinfo{year}{2019}.
\newblock \bibinfo{title}{Integrating gene regulatory pathways into
  differential network analysis of gene expression data}.
\newblock \bibinfo{journal}{Scientific reports} \bibinfo{volume}{9},
  \bibinfo{pages}{5479}.
\bibitem[{Guo et~al.(2020)Guo, Yao, Cao, Liu and Sun}]{Guo2020Shared}
\bibinfo{author}{Guo, X.}, \bibinfo{author}{Yao, D.}, \bibinfo{author}{Cao,
  Q.}, \bibinfo{author}{Liu, L.}, \bibinfo{author}{Sun, L.},
  \bibinfo{year}{2020}.
\newblock \bibinfo{title}{Shared and distinct resting functional connectivity
  in children and adults with attention-deficit/hyperactivity disorder}.
\newblock \bibinfo{journal}{Translational Psychiatry} \bibinfo{volume}{10},
  \bibinfo{pages}{65}.
\bibitem[{He et~al.(2018)He, Ji, Xie et~al.}]{He2018a}
\bibinfo{author}{He, Y.}, \bibinfo{author}{Ji, J.}, \bibinfo{author}{Xie, L.},
  et~al., \bibinfo{year}{2018}.
\newblock \bibinfo{title}{A new insight into underlying disease mechanism
  through semi-parametric latent differential network model}.
\newblock \bibinfo{journal}{BMC Bioinformatics} \bibinfo{volume}{19},
  \bibinfo{pages}{493}.
\bibitem[{Ji et~al.(2017)Ji, He, Feng et~al.}]{ji2017jdinac}
\bibinfo{author}{Ji, J.}, \bibinfo{author}{He, D.}, \bibinfo{author}{Feng, Y.},
  et~al., \bibinfo{year}{2017}.
\newblock \bibinfo{title}{\text{JDINAC}: joint density-based non-parametric
  differential interaction network analysis and classification using
  high-dimensional sparse omics data}.
\newblock \bibinfo{journal}{Bioinformatics} \bibinfo{volume}{33},
  \bibinfo{pages}{3080--3087}.
\bibitem[{Ji et~al.(2020)Ji, He, Liu and Xie}]{Ji2020Brain}
\bibinfo{author}{Ji, J.}, \bibinfo{author}{He, Y.}, \bibinfo{author}{Liu, L.},
  \bibinfo{author}{Xie, L.}, \bibinfo{year}{2020}.
\newblock \bibinfo{title}{Brain connectivity alteration detection via
  matrix-variate differential network model}.
\newblock \bibinfo{journal}{Biometrics, in press} .
\bibitem[{Konrad and Eickhoff(2010)}]{Konrad2010Is}
\bibinfo{author}{Konrad, K.}, \bibinfo{author}{Eickhoff, S.B.},
  \bibinfo{year}{2010}.
\newblock \bibinfo{title}{Is the adhd brain wired differently? a review on
  structural and functional connectivity in attention deficit hyperactivity
  disorder}.
\newblock \bibinfo{journal}{Human Brain Mapping} \bibinfo{volume}{31},
  \bibinfo{pages}{904--916}.
\bibitem[{Leng and Tang(2012)}]{Chenlei2012Sparse}
\bibinfo{author}{Leng, C.}, \bibinfo{author}{Tang, C.Y.}, \bibinfo{year}{2012}.
\newblock \bibinfo{title}{Sparse matrix graphical models}.
\newblock \bibinfo{journal}{Journal of the American Statistical Association}
  \bibinfo{volume}{107}, \bibinfo{pages}{1187--1200}.
\bibitem[{Liu et~al.(2013)Liu, Huang and Ma}]{Liu2013Integrative}
\bibinfo{author}{Liu, J.}, \bibinfo{author}{Huang, J.}, \bibinfo{author}{Ma,
  S.}, \bibinfo{year}{2013}.
\newblock \bibinfo{title}{Integrative analysis of multiple cancer genomic
  datasets under the heterogeneity model}.
\newblock \bibinfo{journal}{Statistics in Medicine} \bibinfo{volume}{32},
  \bibinfo{pages}{3509--3521}.
\bibitem[{Manjari and Allen(2016)}]{Manjari2016Mixed}
\bibinfo{author}{Manjari, N.}, \bibinfo{author}{Allen, G.I.},
  \bibinfo{year}{2016}.
\newblock \bibinfo{title}{Mixed effects models for resampled network statistics
  improves statistical power to find differences in multi-subject functional
  connectivity}.
\newblock \bibinfo{journal}{Frontiers in Neuroence} \bibinfo{volume}{10},
  \bibinfo{pages}{108}.
\bibitem[{Murias et~al.(2006)Murias, Swanson and
  Srinivasan}]{Murias2006Functional}
\bibinfo{author}{Murias, M.}, \bibinfo{author}{Swanson, J.M.},
  \bibinfo{author}{Srinivasan, R.}, \bibinfo{year}{2006}.
\newblock \bibinfo{title}{{Functional Connectivity of Frontal Cortex in Healthy
  and ADHD Children Reflected in EEG Coherence}}.
\newblock \bibinfo{journal}{Cerebral Cortex} \bibinfo{volume}{17},
  \bibinfo{pages}{1788--1799}.
\newblock \DOIprefix\doi{10.1093/cercor/bhl089}.
\bibitem[{Ou~Yang et~al.(2019)Ou~Yang, Zhang, Zhao, Wang, Wang, Lei and
  Yan}]{Ou2018Joint}
\bibinfo{author}{Ou~Yang, L.}, \bibinfo{author}{Zhang, X.F.},
  \bibinfo{author}{Zhao, X.M.}, \bibinfo{author}{Wang, D.D.},
  \bibinfo{author}{Wang, F.L.}, \bibinfo{author}{Lei, B.},
  \bibinfo{author}{Yan, H.}, \bibinfo{year}{2019}.
\newblock \bibinfo{title}{Joint learning of multiple differential networks with
  latent variables}.
\newblock \bibinfo{journal}{IEEE Transactions on Cybernetics}
  \bibinfo{volume}{49}, \bibinfo{pages}{3494--3506}.
\bibitem[{Peng et~al.(2009)Peng, Wang, Zhou et~al.}]{Peng2009Partial}
\bibinfo{author}{Peng, J.}, \bibinfo{author}{Wang, P.}, \bibinfo{author}{Zhou,
  N.}, et~al., \bibinfo{year}{2009}.
\newblock \bibinfo{title}{Partial correlation estimation by joint sparse
  regression models}.
\newblock \bibinfo{journal}{Publications of the American Statistical
  Association} \bibinfo{volume}{104}, \bibinfo{pages}{735--746}.
\bibitem[{Qiu and Zhou(2021)}]{Qiu2021Inference}
\bibinfo{author}{Qiu, Y.}, \bibinfo{author}{Zhou, X.H.}, \bibinfo{year}{2021}.
\newblock \bibinfo{title}{Inference on multi-level partial correlations based
  on multi-subject time series data}.
\newblock \bibinfo{journal}{Journal of the American Statistical Association, in
  press} .
\bibitem[{Smith(2012)}]{Smith2012The}
\bibinfo{author}{Smith, S.M.}, \bibinfo{year}{2012}.
\newblock \bibinfo{title}{The future of \text{FMRI} connectivity}.
\newblock \bibinfo{journal}{Neuroimage} \bibinfo{volume}{62},
  \bibinfo{pages}{1257--1266}.
\newblock \DOIprefix\doi{10.1016/j.neuroimage.2012.01.022}.
\bibitem[{Stoodley(2016)}]{Stoodley2016cerebellum}
\bibinfo{author}{Stoodley, C.J.}, \bibinfo{year}{2016}.
\newblock \bibinfo{title}{The cerebellum and neurodevelopmental disorders}.
\newblock \bibinfo{journal}{Cerebellum (London, England)} \bibinfo{volume}{15},
  \bibinfo{pages}{34--37}.
\newblock \URLprefix \url{https://pubmed.ncbi.nlm.nih.gov/26298473}.
\bibitem[{Tian et~al.(2016)Tian, Gu and Ma}]{tian2016identifying}
\bibinfo{author}{Tian, D.}, \bibinfo{author}{Gu, Q.}, \bibinfo{author}{Ma, J.},
  \bibinfo{year}{2016}.
\newblock \bibinfo{title}{Identifying gene regulatory network rewiring using
  latent differential graphical models}.
\newblock \bibinfo{journal}{Nucleic acids research} \bibinfo{volume}{44},
  \bibinfo{pages}{e140--e140}.
\bibitem[{Vaidya(2012)}]{vaidya2012neuro}
\bibinfo{author}{Vaidya, C.J.}, \bibinfo{year}{2012}.
\newblock \bibinfo{title}{Neurodevelopmental abnormalities in adhd}.
\newblock \bibinfo{journal}{Current topics in behavioral neurosciences}
  \bibinfo{volume}{9}, \bibinfo{pages}{49--66}.
\newblock \DOIprefix\doi{10.1007/7854{\_}2011{\_}138}.
\bibitem[{Valera et~al.(2007)Valera, Faraone, Murray and
  Seidman}]{Valera2007meta}
\bibinfo{author}{Valera, E.M.}, \bibinfo{author}{Faraone, S.V.},
  \bibinfo{author}{Murray, K.E.}, \bibinfo{author}{Seidman, L.J.},
  \bibinfo{year}{2007}.
\newblock \bibinfo{title}{Meta-analysis of structural imaging findings in
  attention-deficit/hyperactivity disorder}.
\newblock \bibinfo{journal}{Biological Psychiatry} \bibinfo{volume}{61},
  \bibinfo{pages}{1361--1369}.
\newblock \DOIprefix\doi{10.1016/j.biopsych.2006.06.011}.
\bibitem[{Wang et~al.(2016)Wang, Jian, B. and Ying}]{WangAn2016}
\bibinfo{author}{Wang, Y.}, \bibinfo{author}{Jian, K.}, \bibinfo{author}{B.,
  K.P.}, \bibinfo{author}{Ying, G.}, \bibinfo{year}{2016}.
\newblock \bibinfo{title}{An efficient and reliable statistical method for
  estimating functional connectivity in large scale brain networks using
  partial correlation}.
\newblock \bibinfo{journal}{Frontiers in Neuroscience} \bibinfo{volume}{10},
  \bibinfo{pages}{123}.
\newblock \DOIprefix\doi{10.3389/fnins.2016.00123}.
\bibitem[{the ADHD-200~dataset website(2019)}]{neurobureau.projects.nitrc.org}
\bibinfo{author}{the ADHD-200~dataset website}, \bibinfo{year}{2019}.
\newblock \bibinfo{title}{The neuro bureau}.
\newblock
  \bibinfo{howpublished}{\url{http://neurobureau.projects.nitrc.org/ADHD200/Data.html}}.
\newblock \bibinfo{note}{(Accessed March 7, 2019)}.
\bibitem[{Wieringen and Peeters(2016)}]{Wieringen2016Ridge}
\bibinfo{author}{Wieringen, W.N.V.}, \bibinfo{author}{Peeters, C.F.W.},
  \bibinfo{year}{2016}.
\newblock \bibinfo{title}{Ridge estimation of inverse covariance matrices from
  high-dimensional data}.
\newblock \bibinfo{journal}{Computational Statistics $\&$ Data Analysis}
  \bibinfo{volume}{103}, \bibinfo{pages}{284--303}.
\bibitem[{Xia et~al.(2013)Xia, Wang and He}]{Xia2013BrainNet}
\bibinfo{author}{Xia, M.}, \bibinfo{author}{Wang, J.}, \bibinfo{author}{He,
  Y.}, \bibinfo{year}{2013}.
\newblock \bibinfo{title}{Brainnet viewer: a network visualization tool for
  human brain connectomics}.
\newblock \bibinfo{journal}{PloS one} \bibinfo{volume}{8},
  \bibinfo{pages}{e68910--e68910}.
\newblock \DOIprefix\doi{10.1371/journal.pone.0068910}.
\bibitem[{Xia et~al.(2015)Xia, Cai and Cai}]{Xia2015Testing}
\bibinfo{author}{Xia, Y.}, \bibinfo{author}{Cai, T.}, \bibinfo{author}{Cai,
  T.T.}, \bibinfo{year}{2015}.
\newblock \bibinfo{title}{Testing differential networks with applications to
  detecting gene-by-gene interactions}.
\newblock \bibinfo{journal}{Biometrika} \bibinfo{volume}{102},
  \bibinfo{pages}{247--266}.
\bibitem[{Xia and Li(2017)}]{Xia2017Hypothesis}
\bibinfo{author}{Xia, Y.}, \bibinfo{author}{Li, L.}, \bibinfo{year}{2017}.
\newblock \bibinfo{title}{Hypothesis testing of matrix graph model with
  application to brain connectivity analysis}.
\newblock \bibinfo{journal}{Biometrics} \bibinfo{volume}{73},
  \bibinfo{pages}{780--791}.
\bibitem[{Xia and Li(2019)}]{Xia2018matrix}
\bibinfo{author}{Xia, Y.}, \bibinfo{author}{Li, L.}, \bibinfo{year}{2019}.
\newblock \bibinfo{title}{Matrix graph hypothesis testing and application in
  brain connectivity alternation detection}.
\newblock \bibinfo{journal}{Statistica Sinica} \bibinfo{volume}{29},
  \bibinfo{pages}{303--328}.
\bibitem[{Yang et~al.(2014)Yang, Huang and Zhou}]{Yang2014Concave}
\bibinfo{author}{Yang, G.}, \bibinfo{author}{Huang, J.}, \bibinfo{author}{Zhou,
  Y.}, \bibinfo{year}{2014}.
\newblock \bibinfo{title}{Concave group methods for variable selection and
  estimation in high-dimensional varying coefficient models}.
\newblock \bibinfo{journal}{Science China Mathematics} \bibinfo{volume}{57},
  \bibinfo{pages}{2073--2090}.
\bibitem[{Yin and Li(2012)}]{Yin2012Model}
\bibinfo{author}{Yin, J.}, \bibinfo{author}{Li, H.}, \bibinfo{year}{2012}.
\newblock \bibinfo{title}{Model selection and estimation in the matrix normal
  graphical model}.
\newblock \bibinfo{journal}{Journal of Multivariate Analysis}
  \bibinfo{volume}{107}, \bibinfo{pages}{119}.
\bibitem[{Yuan et~al.(2015)Yuan, Xi and Deng}]{Yuan2015Differential}
\bibinfo{author}{Yuan, H.}, \bibinfo{author}{Xi, R.}, \bibinfo{author}{Deng,
  M.}, \bibinfo{year}{2015}.
\newblock \bibinfo{title}{Differential network analysis via the lasso penalized
  d-trace loss}.
\newblock \bibinfo{journal}{Biometrika} \bibinfo{volume}{104},
  \bibinfo{pages}{755--770}.
\bibitem[{Yuan(2010)}]{Yuan2010High}
\bibinfo{author}{Yuan, M.}, \bibinfo{year}{2010}.
\newblock \bibinfo{title}{High dimensional inverse covariance matrix estimation
  via linear programming}.
\newblock \bibinfo{journal}{Journal of Machine Learning Research}
  \bibinfo{volume}{11}, \bibinfo{pages}{2261--2286}.
\bibitem[{Zhang(2010)}]{Zhang2010Nearly}
\bibinfo{author}{Zhang, C.H.}, \bibinfo{year}{2010}.
\newblock \bibinfo{title}{Nearly unbiased variable selection under minimax
  concave penalty}.
\newblock \bibinfo{journal}{Annals of Stats} \bibinfo{volume}{38},
  \bibinfo{pages}{894--942}.
\bibitem[{Zhang et~al.(2020)Zhang, Sun and Li}]{Zhang2020Mixed}
\bibinfo{author}{Zhang, J.}, \bibinfo{author}{Sun, W.W.}, \bibinfo{author}{Li,
  L.}, \bibinfo{year}{2020}.
\newblock \bibinfo{title}{Mixed-effect time-varying network model and
  application in brain connectivity analysis}.
\newblock \bibinfo{journal}{Journal of the American Statistical Association}
  \bibinfo{volume}{115}, \bibinfo{pages}{2022--2036}.
\bibitem[{Zhang et~al.(2017)Zhang, Ou-Yang and Yan}]{Zhang2017Incorporating}
\bibinfo{author}{Zhang, X.F.}, \bibinfo{author}{Ou-Yang, L.},
  \bibinfo{author}{Yan, H.}, \bibinfo{year}{2017}.
\newblock \bibinfo{title}{Incorporating prior information into differential
  network analysis using nonparanormal graphical models}.
\newblock \bibinfo{journal}{Bioinformatics} \bibinfo{volume}{33},
  \bibinfo{pages}{2436}.
\bibitem[{Zhao et~al.(2014)Zhao, Cai and Li}]{zhao2014direct}
\bibinfo{author}{Zhao, S.D.}, \bibinfo{author}{Cai, T.T.}, \bibinfo{author}{Li,
  H.}, \bibinfo{year}{2014}.
\newblock \bibinfo{title}{Direct estimation of differential networks}.
\newblock \bibinfo{journal}{Biometrika} \bibinfo{volume}{101},
  \bibinfo{pages}{253--268}.
\bibitem[{Zhao et~al.(2012)Zhao, Liu, Roeder, Lafferty and
  Wasserman}]{zhao2012huge}
\bibinfo{author}{Zhao, T.}, \bibinfo{author}{Liu, H.}, \bibinfo{author}{Roeder,
  K.}, \bibinfo{author}{Lafferty, J.}, \bibinfo{author}{Wasserman, L.},
  \bibinfo{year}{2012}.
\newblock \bibinfo{title}{The huge package for high-dimensional undirected
  graph estimation in r}.
\newblock \bibinfo{journal}{Journal of Machine Learning Research}
  \bibinfo{volume}{13}, \bibinfo{pages}{1059--1062}.
\bibitem[{Zhou(2014)}]{Zhou2014Gemini}
\bibinfo{author}{Zhou, S.}, \bibinfo{year}{2014}.
\newblock \bibinfo{title}{Gemini: Graph estimation with matrix variate normal
  instances}.
\newblock \bibinfo{journal}{Annals of Statistics} \bibinfo{volume}{42},
  \bibinfo{pages}{532--562}.
\bibitem[{Zhu and Li(2018)}]{Zhu2018multiple}
\bibinfo{author}{Zhu, Y.}, \bibinfo{author}{Li, L.}, \bibinfo{year}{2018}.
\newblock \bibinfo{title}{Multiple matrix gaussian graphs estimation}.
\newblock \bibinfo{journal}{Journal of the Royal Statistical Society: Series B
  (Statistical Methodology)} \bibinfo{volume}{80}, \bibinfo{pages}{927--950}.

\end{thebibliography}


%






\end{document}